\documentclass[conference,a4paper]{IEEEtran}

\usepackage{multirow}
\usepackage[algoruled, linesnumbered]{algorithm2e}
\usepackage{graphicx}
\usepackage{psfrag}
\usepackage{epsfig}
\usepackage{subfigure}
\usepackage{amsmath}
\usepackage{array}
\usepackage{amssymb}
\usepackage{graphicx}
\usepackage{psfrag}
\usepackage{epsfig}
\usepackage{subfigure}
\usepackage{url}
\usepackage{stfloats}
\usepackage{multirow}
\usepackage{array}
\usepackage{slashbox}
\usepackage{array}
\usepackage{epstopdf}

\IEEEoverridecommandlockouts
\IEEEpubid{\makebox[\columnwidth]{978--1--4673--6540--6/15/\$31.00~\copyright~2015~IEEE \hfill} \hspace{\columnsep}\makebox[\columnwidth]{ }}
\begin{document}

\title{Novel Speech Features for Improved Detection of Spoofing Attacks}

\author{\IEEEauthorblockN{Dipjyoti Paul\IEEEauthorrefmark{1},
Monisankha Pal\IEEEauthorrefmark{2},
Goutam Saha\IEEEauthorrefmark{3}}
\IEEEauthorblockA{Email: \IEEEauthorrefmark{1}dipjyotipaul@ece.iitkgp.ernet.in, \IEEEauthorrefmark{2}monisankhapal@iitkgp.ac.in, \IEEEauthorrefmark{3}gsaha@ece.iitkgp.ernet.in}
\IEEEauthorblockA{\IEEEauthorrefmark{1}\IEEEauthorrefmark{2}\IEEEauthorrefmark{3} Dept of Electronics \& Electrical Communication Engineering, Indian Institute of Technology, Kharagpur, India}}

\maketitle

\begin{abstract}
Now-a-days, speech-based biometric systems such as automatic speaker verification (ASV) are highly prone to spoofing attacks by an imposture. With recent development in various voice conversion (VC) and speech synthesis (SS) algorithms, these spoofing attacks can pose a serious potential threat to the current state-of-the-art ASV systems. To impede such attacks and enhance the security of the ASV systems, the development of efficient anti-spoofing algorithms is essential that can differentiate synthetic or converted speech from natural or human speech. In this paper, we propose a set of novel speech features for detecting spoofing attacks. The proposed features are computed using alternative frequency-warping technique and formant-specific block transformation of filter bank log energies. We have evaluated existing and proposed features against several kinds of synthetic speech data from ASVspoof 2015 corpora. The results show that the proposed techniques outperform existing approaches for various spoofing attack detection task. The techniques investigated in this paper can also accurately classify natural and synthetic speech as equal error rates (EERs) of $0\%$ have been achieved.
\end{abstract}


\begin{figure*}[t]
  \centering

  \includegraphics[width=1\textwidth]{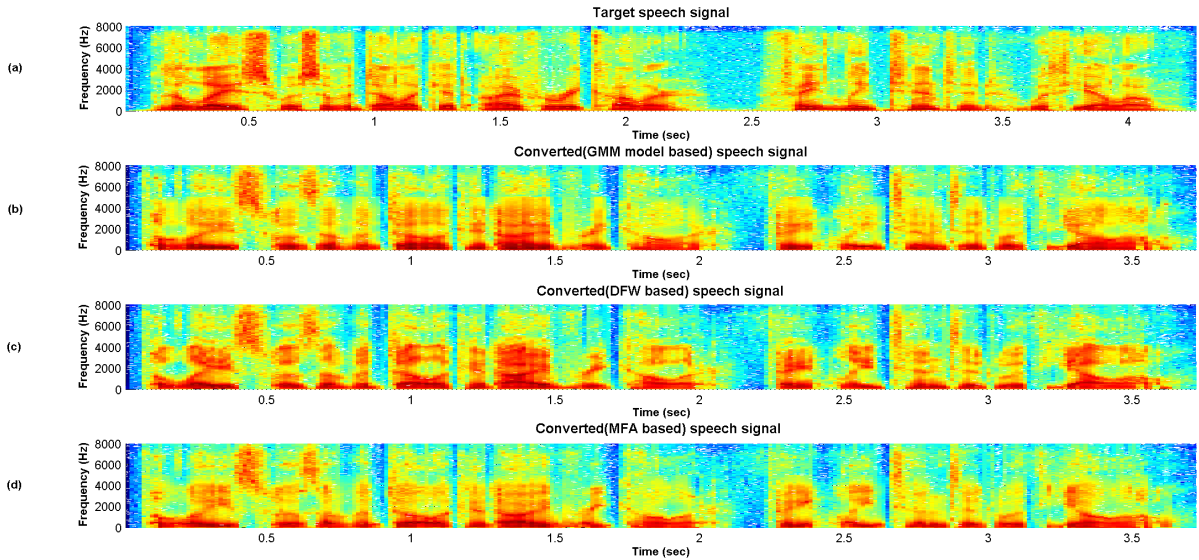}
  \caption{Spectrogram of ``\textit{The lace was of a delicate ivory color, faintly tinted with yellow}'' from CMU-ARCTIC database of natural and different converted speech signals. The converted signals are generated by using VC techniques based on (b) GMM, (c) dynamic frequency warping (DFW), and (d) mixture of factor analyzer (MFA) based algorithms, respectively.}
\label{spec_plot}
\end{figure*}

\section{Introduction}\label{Section:Introduction}
Synthetic speech signal can be obtained non-intrusively by using different voice conversion (VC) \cite{stylianou1998continuous} and speech synthesis (SS) \cite{moulines1990pitch} techniques. Due to the advancements in speech technology, VC and SS techniques have been developed a lot which can be used to generate synthetic voice with excellent quality and naturalness like a real human speech. These techniques generally concentrate on mapping the spectral characteristics. Among the VC techniques, Gaussian mixture model (GMM) based \cite{kain1998spectral}, frequency warping based \cite{toda2001voice} and unit-selection based \cite{sundermann2006text} approaches are well known. Now-a-days, speech synthesis technique like hidden Markov model (HMM) based text-to-speech (TTS) \cite{yoshimuray1999simultaneous} and speaker adapted speech synthesis \cite{tamura1998speaker} are very much popular. These synthetic speechs can be deliberately used as a spoofing attack to deceive the speech-based biometric systems like ASV systems \cite{kinnunen2012vulnerability}. Thus in order to enhance the security of ASV systems, development of an efficient anti-spoofing  mechanism that can distinguish between real speech and synthetic speech is highly essential.

During the last decade, a significant effort has been made to develop the countermeasures that can prevent spoofing attack in ASV systems \cite{wu2015spoofing}. An anti-spoofing system consists of two major blocks: feature extraction or front-end for speech parameterization and classifier or back-end for modeling real and synthetic speech class. In feature level, several features such as  mel-frequency cepstrum coefficient (MFCC), modified group delay function and cos-phase features were studied for anti-spoofing \cite{wu2012detecting}. Phase information is usually neglected during the process of speech synthesis. Therefore, the phase based features can be used as an informative cue for synthetic speech detection. Phase based features like relative phase shift (RPS) \cite{de2012evaluation, sanchez2015toward}, modified group delay phase \cite{wu2012detecting, wu2012study} were investigated for reliable detection of HMM-based synthetic speech. Modulation features extracted from magnitude as well as phase spectrum were also used to detect HMM-based synthetic speech from natural speech \cite{wu2013synthetic}. Better results were reported in score level fusion when phase based modulation and short-term spectral features were used. Prosodic features using several pitch statistics were also used for the anti-spoofing task \cite{de2012synthetic}. Other countermeasure techniques developed for this purpose were based on high level dynamic features and voice quality assessment \cite{alegre2012spoofing}, pairwise distance between input and target speaker data in training \cite{alegre2013spoofing} and local binary pattern analysis \cite{alegre2013one, alegre2013new}. Moreover, the robustness of various state-of-the-art speech based biometric systems against different voice conversion attacks and a possible solution to prevent such attacks were provided in \cite{pal2015robustness}. On the other hand, a useful comparative study showing vulnerability of various types of short-term, dynamic and complementary features for spoofing detection system was given in \cite{sahidullah2015comparison}. Different classifier level experiments were also conducted in \cite{hanilcci2015classifiers} where GMM trained using maximum-likelihood (GMM-ML) criterion [18] classifier outperforms other methods for different kinds of spoofing attacks.

In this paper, we have evaluated several features for spoofing detection which were successfully used in speaker verification task \cite{kinnunen2010overview}. Usually voice conversion techniques utilize mel-warping where lower frequency components are more emphasized. As a result, the higher frequency regions are not well mapped. This is also illustrated in the spectrogram in Fig. \ref{spec_plot} for natural and three different voice converted speech signals. In order to represent high frequency speech information for spoofing detection task, we propose features based on inverted warping scale. In cepstral feature extraction algorithm, the filter bank structure is computed in such a way that it mimics human auditory perception. Moreover, most of the cepstral feature coefficients are determined by giving more emphasis on low frequency regions due to the mel-warped filter bank structure. Here, to get more information from high frequency region, the inverted warping scale is used for computing filter bank structure. In speaker recognition task \cite{campbell1997speaker}, it is found that speech-signal-based frequency warping (SFCC) and mel overlapped block transform (MOBT) \cite{sahidullah2012design} based short-term spectral features provide better performance than standard MFCC. Motivated by this, we formulated inverted version of those cepstral coefficients for spoofing attack detection. Experiments are conducted on the development section of the recently released ASVspoof $2015$ speech corpus. We have obtained improved classification accuracy with our proposed features for five different VC and SS techniques.

The remainder of the paper is organized as follows: different feature extraction techniques are described in Section \ref{Section:FEATURE EXTRACTION}. Section \ref{Section:Experimental Setup} provides database description and experimental framework. In Section \ref{Section:Results and Discussion}, experimental results are discussed. The conclusion of the work is summarized in Section \ref{Section:Conclusion}.

\section{\lowercase{FEATURE EXTRACTION TECHNIQUES FOR COUNTERMEASURES}}\label{Section:FEATURE EXTRACTION}
\begin{figure*}[t]
  \centering

  \includegraphics[width=1\textwidth]{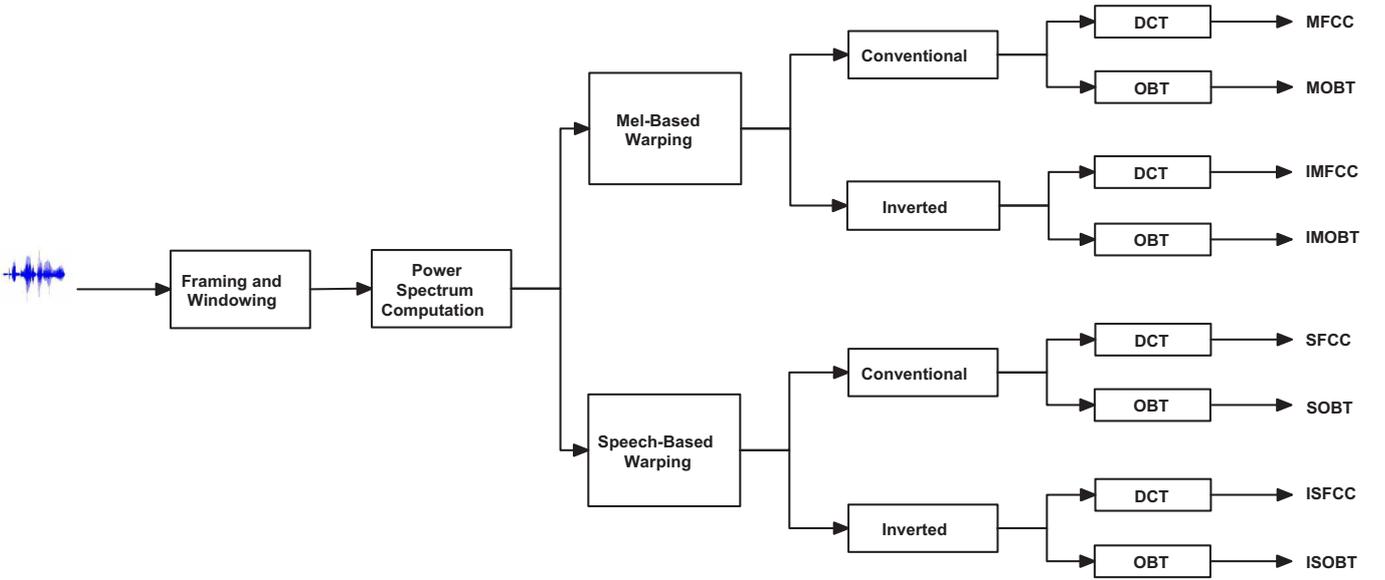}
  \caption{The basic functional block diagram of different feature extraction techniques used in our work.}
\label{blockdiagram}
\end{figure*}

Real and synthetic speech can be discriminated by their spectral characteristics which is very much prominent in Fig. \ref{spec_plot}. It is also clear that the discriminative information between human and synthetic speech signal resides in the high frequency zone of their corresponding spectrograms. Therefore, feature related information captured from the high frequency region can be employed for our classification task. Several short-term spectral feature extraction techniques are analyzed to detect spoofing attacks in \cite{sahidullah2015comparison}. The detailed descriptions of all the existing and proposed feature extraction algorithms are  described in this section. The basic functional block diagram of their extraction process is pictorially represented in Fig. \ref{blockdiagram}.

\subsection{Mel-Frequency Cepstral Coefficient (MFCC)}
MFCC \cite{davis1980comparison,apoddar15} feature captures spectral and phonetic information related to speech signal. A brief description of this feature extraction technique can be described as follows:

Discrete Fourier transform (DFT) is computed on the framed speech signal  $s(n)$ to estimate the short-term power spectrum. Each speech frame is passed through a mel scaled triangular filter bank, where the output would be the multiplication of frequency response of the framed speech signal and the filters used in the filter bank. The mel scale can be represented as $f_{mel}=2595\log_{10}\left(1+\frac{f}{700}\right)$, where $f$ is frequency in Hz. Mel filter bank log energy (MFLE) of speech frames are derived using the logarithm operation to the filterbank energies. At the final stage, discrete cosine transform (DCT) is applied to all the MFLE coefficients to generate decorrelated feature vectors.

The inverted mel-frequency scale of the competing filter bank structure is used to extract inverse mel-frequency cepstral coefficient (IMFCC) that can capture information in the higher frequency components \cite{chakroborty2007improved,SandiThesis2008}.

\subsection{Mel-warped Overlapped Block Transformation (MOBT)}
In block-based transformation \cite{sahidullah2012design}, the MFLE coefficients of each speech frame are divided into non-overlapping and overlapping blocks. In order to compute cepstral features from MFLE using block transformation, linear DCT kernel is applied to process each block individually. This kernel is selected for transforming segments of MFLE coefficients into cepstral feature vectors.

MOBT \cite{SahidThesis} feature is calculated in the same manner as MFCC, except the fact that DCT on the whole MFLE coefficients is replaced by block-based DCT. Here, each block corresponds to a chunk of filter bank log energies. The first and the second block in MOBT computation are chosen in such a manner so that they can cover the first ($F1$) and second ($F2$), third ($F3$) formant frequencies respectively \cite{o1987speech}. After applying the block-based transformation kernel, a 22-dimensional feature vector is obtained from each speech frame.

The inverted mel scale can also be used for the purpose of extracting meaningful information from the higher frequency components of the speech signal.
\subsection{Speech-Signal-Based Frequency Cepstral Coefficient (SFCC)}
SFCC feature is obtained using speech-signal-based frequency warping technique \cite{paliwal2009speech, sarangi2012novel}. The processing steps of this method are described as follows.

Short-term Fourier transform (STFT) is computed from the given input speech signal $x(n)$. After that, periodogram-based power spectral density (PSD) is estimated for each frame of the speech signal. It can be written as
\begin{equation}
  S(n,w) = \frac{1}{N}|X(n,w)|^{2},
\end{equation}

where $\textit{N}$ is the number of samples in analysis window. By averaging $S(n,w)$ over the entire speech corpus, ensemble energy $\bar{S}(w)$ is calculated. A logarithm of the ensemble average spectrum is calculated and divided into equal area, such that
\begin{equation}
  A_{i}=\int_{w_{i}}^{w_{i+1}}\log\bar{S}(w)dw, \hspace{10pt} i=1, \ldots , M,
\end{equation}
and
\begin{equation}
  A_{i}=A_{i+1},\hspace{10pt} i=1, \ldots , M-1,
\end{equation}
where $A_{i}$ is the $\textit{i}$-th interval area. $w_{i}$ and $w_{i+1}$ denotes lower and upper cut-off frequencies respectively. $M$ point speech-signal-based frequency warping function can be calculated as
\begin{equation}
  W\left(\frac{w_{i}+w_{i+1}}{2}\right)=\frac{i}{M},  \hspace{10pt} i=1, \ldots , M,
\end{equation}
where $W(w)$ function becomes continuous as $M$ approaches to infinity and its value lies between $1$ and $-1$.

Spectral domain to cepstral domain feature calculation is similar to the procedure followed in MFCC computation. The difference is that instead of using mel scale filter bank, the speech-signal-based warping function is used to generate triangular filter bank. The inverted warping function based filter bank can also be used for ISFCC feature computation.

Motivated by the computation of MOBT, SFCC can also be computed alternatively in reference to the block-based computation of feature. The proposed feature SOBT is the combination of SFCC and MOBT features. In this feature, the scale of the filter bank is generated in a speech-signal-based adaptive manner. The inverted version of this feature is also introduced, which can be called as ISOBT.



\section{Experimental Setup}\label{Section:Experimental Setup}
\subsection{Speech Corpora}
The experiment for the performance evaluation of several anti-spoofing methods has been conducted on recently released ASVspoof $2015$ corpus whose detailed description is available in \cite{wu10asvspoof}. The synthetic speech is generated by five different techniques as follows: frame selection (FS) based VC algorithm (S1), first mel-cepstral (C1) based VC algorithm (S2), HMM-based synthesis algorithm (S3 and S4) and lastly, VC toolkit within Festvox system (S5). The database contains human and spoofed speech data for both the training and development set. Training data consists of $3750$ utterances for human and 12625 for spoofed speech signals. Development data set contains 3497 utterances for genuine and 49875 utterances for synthetic speech.
\subsection{Parameters of Feature Extraction}

Features are extracted from speech frames with frame size 20 ms and of overlap 50\%. Windowing is performed using Hamming window \cite{sahidullah2013novel}. Voice activity detector (VAD) \cite{sahidullah2012comparison} is not used for all the experiments performed. We have also included the energy coefficients in speech feature computation. For all the experiments, 20-dimensional short-term features (MFCC, MOBT and SFCC) vectors are used along with their delta (20) and double delta (20) coefficients. Static, static and dynamic and only dynamic feature elements are incorporated for anti-spoofing operation. Dynamic feature is calculated using three consecutive speech frames. On the other hand in MOBT, two blocks are set to contain seven and fifteen number of filters. Static, static+$\Delta\Delta^{2}$ and only $\Delta\Delta^{2}$ feature vectors of dimension 22, 66 an 44 are considered respectively for overlapped block transformation features. The frequency bands covered by the two segments are $0-1128$ Hz and $575-8000$ Hz respectively. They are dominated by the first three formants $(F1)$ and $(F2)$, $(F3)$. The filter bank of the SFCC and ISFCC features are generated using training data set of the database and after that cepstral features are computed.

\subsection{Classifier}
Proposed and existing features are evaluated using GMM-ML classifier since it is widely used for speaker verification systems. Ten iterations of expectation-maximization (EM) \cite{reynolds1995robust} algorithm are used to estimate each class parameters via maximum-likelihood criteria. The number of Gaussian components is set as 512. Two target models $\boldsymbol{\lambda}_{natural}$ and $\boldsymbol{\lambda}_{synthetic}$  are created from natural and synthetic speech data respectively. The log-likelihood score is calculated as,

\begin{equation}
\Lambda(\mathbf{X})=\mathcal{L}(\textbf{X}|\boldsymbol{\lambda}_{natural})-\mathcal{L}(\mathbf{X}|\boldsymbol{\lambda}_{synthetic}),
\end{equation}
where $\mathbf{X}=\{\mathbf{x}_{1},\ldots, \mathbf{x}_{T}\}$ is the feature matrix of the test utterance, $T$ is the number of frames and $\mathcal{L} (\textbf{X}|\boldsymbol{\lambda})=(1/T)\sum_{t=1}^{T}\log p(\textbf{x}_{t}|\boldsymbol{\lambda})$ is the average log-likelihood of \textbf{X} given GMM model $\boldsymbol{\lambda}$. In our experiment, all the training data is used for modeling and development data is used for testing purpose.

\subsection{Performance Evaluation Metric}
Equal error rate (EER) is used as the performance metric to evaluate the spoofing attack detection. We use Bosaris toolkit \footnote{{https://sites.google.com/site/bosaristoolkit/}} to calculate the EER using receiver operating characteristics convex hull (ROCCH) method. The lower the value of EER, better is the anti-spoofing performance. The average and individual EER are reported for all kind of VC and SS algorithms \cite{wu2014aasvspoof}.

\begin{table}[h!]
\renewcommand{\arraystretch}{1.6}
\centering \caption{Comparative accuracy (EER in \%) of existing and proposed cepstral features for GMM-ML classifier in ASVspoof 2015 corpora}
\setlength{\tabcolsep}{5pt}
\label{my-label}
\begin{tabular}{|c|c|c|c|c|c|c|c|}
\hline \multicolumn{1}{|l|}{Feature}          &\multicolumn{1}{c|}{Type}          &  S1   & S2     & S3  & S4  &  S5    &  Avg. \\ \hline
\multirow{3}{*}{MFCC}                  & Static                        & 0.981  & 11.720  &  0.000  &  0.000  &  6.030  &  3.746    \\ \cline{2-8}
                                       & Static+$\Delta \Delta^{2}$    & 0.036  & 4.597   &  0.000  &  0.000  &  0.649  &  1.056    \\ \cline{2-8}
                                       & $\Delta \Delta^{2}$           & 0.037  & 0.657   &  0.000  &  0.000  &  0.020  &  0.143    \\ \hline \hline

\multirow{3}{*}{MOBT}            & \multicolumn{1}{c|}{Static}         & 0.897  & 10.451  &  0.000  &  0.000  &  4.714  &  \textbf{3.212}    \\ \cline{2-8}
                                       & Static+$\Delta \Delta^{2}$    & 0.016  & 3.290   &  0.000  &  0.000  &  0.349  &  \textbf{0.731}    \\ \cline{2-8}
                                       & $\Delta \Delta^{2}$           & 0.016  & 0.455   &  0.000  &  0.000  &  0.017  &  \textbf{0.098}    \\ \hline \hline

\multirow{3}{*}{SFCC}                  & \multicolumn{1}{c|}{Static}   & 2.395  & 18.402   &  0.000  & 0.000   &  5.750  & 5.309     \\ \cline{2-8}
                                       & Static+$\Delta \Delta^{2}$    & 0.025  & 7.718   &  0.000  & 0.000   &  0.582  & 1.665     \\ \cline{2-8}
                                       & $\Delta \Delta^{2}$           & 0.062  & 2.205   &  0.000  & 0.000   &  0.077  & 0.469     \\ \hline \hline

\multirow{3}{*}{SOBT}            & \multicolumn{1}{c|}{Static}         & 2.360  &  16.664  &  0.000  &  0.000   & 5.851  & 4.975    \\ \cline{2-8}
                                       & Static+$\Delta \Delta^{2}$    & 0.037   &  6.038  & 0.000  &  0.000   & 0.326  &  1.280    \\ \cline{2-8}
                                       & $\Delta \Delta^{2}$           & 0.053   & 1.555  & 0.000 & 0.000 & 0.154  &  0.352    \\ \hline
\end{tabular}
\label{table: 1}
\end{table}

\begin{table}[h!]
\renewcommand{\arraystretch}{1.6}
\centering \caption{Comparative accuracy (EER in \%) of existing and proposed inverted cepstral features for GMM-ML classifier in ASVspoof 2015 corpora}
\setlength{\tabcolsep}{5pt}
\label{my-label}
\begin{tabular}{|c|c|c|c|c|c|c|c|}
\hline \multicolumn{1}{|l|}{Feature}          &\multicolumn{1}{c|}{Type}     &  S1   & S2     & S3  & S4  &  S5    &  Avg. \\ \hline

\multirow{3}{*}{IMFCC}                 & \multicolumn{1}{c|}{Static}                & 0.142  & 4.777   &  0.000  &  0.000  &  3.215  &  1.627    \\ \cline{2-8}
                                       & Static+$\Delta \Delta^{2}$                 & 0.017  & 1.749   &  0.000  &  0.000  &  0.252  &  0.404    \\ \cline{2-8}
                                       & $\Delta \Delta^{2}$                        & 0.030  & 0.141   &  0.039  &  0.057  &  0.000   &  0.042   \\ \hline \hline

\multirow{3}{*}{IMOBT} & \multicolumn{1}{c|}{Static}                                & 0.000     & 0.290   &  0.000  &  0.000   &  1.673  & 0.393     \\ \cline{2-8}
                                       & Static+$\Delta \Delta^{2}$                 & 0.000     & 0.078   &  0.000  &  0.000   &  0.047  & 0.025     \\ \cline{2-8}
                                       & $\Delta \Delta^{2}$                        & 0.000     & 0.000   &  0.000  &  0.000   &  0.000      &  \textbf{0.000}   \\ \hline \hline

\multirow{3}{*}{ISFCC}       & \multicolumn{1}{c|}{Static}                          & 0.037  & 1.585   &  0.000  & 0.000    & 0.835  & 0.491     \\ \cline{2-8}
                                       & Static+$\Delta \Delta^{2}$                 &  0.000    & 0.587   &  0.000  & 0.000    & 0.089  & 0.135     \\ \cline{2-8}
                                       & $\Delta \Delta^{2}$                        &  0.000    & 0.107   &  0.037 & 0.045 & 0.024  & 0.043     \\ \hline \hline

\multirow{3}{*}{ISOBT}            & \multicolumn{1}{c|}{Static}                     & 0.000  & 0.104   & 0.000   &  0.000   & 0.399  &  \textbf{0.101}   \\ \cline{2-8}
                                       & Static+$\Delta \Delta^{2}$                 &  0.000   &  0.009  & 0.000  &  0.000   & 0.010  &   \textbf{0.004}   \\ \cline{2-8}
                                       & $\Delta \Delta^{2}$                        &  0.000   & 0.000  & 0.000 & 0.000 & 0.000  &  \textbf{0.000}    \\ \hline
\end{tabular}
\label{table: 2}
\end{table}
\section{Results and Discussion}\label{Section:Results and Discussion}
The performance evaluation of several short-term spectral features in the anti-spoofing systems for different synthetic data is shown in Table \ref{table: 1}. The results are reported for various feature combinations like static, static+$\Delta\Delta^{2}$ and $\Delta\Delta^{2}$. Table \ref{table: 1} shows that both the S3 and S4 synthetic data are easy to detect than all other different features. We observe that the performance of the spoofing detector to distinguish between natural and synthetic speech data increases with addition of $\Delta\Delta^{2}$ to static features. The table also illustrates that amongst the four different features, the MOBT feature yields less EER value than other features. This can be probably explained by the fact that MOBT features extract formant specific speech information more efficiently, where the discriminative information resides. The table also shows that our GMM-ML recognizer with all the features faces difficulty in classification for S2 converted data.

In order to further reduce the EER value of the anti-spoofing system, we also incorporated the inverted version of the above mentioned features. The results are shown in Table \ref{table: 2}. The table shows that features in inverted scale further reduce the EER value for all sets of synthetic data. We also find that the newly formulated IMOBT and ISFCC features outperform existing IMFCC. The best result is obtained for our proposed feature, i.e. ISOBT. It produces zero EER value or 100\% accuracy in classification for $\Delta\Delta^{2}$ case which is better than any results given in Table \ref{table: 2} \cite{wu10asvspoof} . The usefulness of the inverted feature over static and dynamic features in our classification task can be justified by the fact that discriminative information between natural and synthetic data lies in the high-frequency region of the spectrum. Usually in all the speech synthesis and voice conversion techniques, the high frequency components of the speech signal are not considered properly. The more visible formant structure in the high-frequency component in the natural speech signal is distorted in the synthetic speech signal. In inverted feature extraction, the filter bank is applied on the flipped version of the spectrum. Therefore, it captures high-frequency information more accurately.

\section{Conclusion}\label{Section:Conclusion}
In this paper, we have investigated features for efficiently detecting speech-based spoofing attacks. The important findings of our study can be summarized as follows:
\begin{itemize}
\item \textbf{Importance of high-frequency components:} It is found that high frequency regions are more informative for recognizing synthetic speech. The experimental results show that speech parameterizations using inverted-scale perform better than their conventional formulations.
\item \textbf{Usefulness of dynamic coefficients:} Dynamic features give better performance than static as well as combination features. From these results we can conclude that dynamic component of all the studied (existing and proposed) features carry more discriminative information for the classification of human and synthetic speech.
\item \textbf{Effectiveness of block transformation:} The block-based features are computed using segment wise transformation of log energies unlike full-band DCT. Each block contains formant-specific spectral information. In our study, we find that block-based approach gives better improvement in spoofing detection.
\item \textbf{Improvement with speech-frequency-based frequency warping:} We have also investigated that filter bank generated from speech-signal-based warped frequency scale improves the accuracy of spoofing detection than conventional mel-frequency based computing.
\end{itemize}

Based on the above findings, we have introduced novel speech features which are dynamic coefficients of cepstral features that are computed by formant-specific block transformation of log energies derived using inverted speech-signal-based warping method. Proposed features yield $0\%$ classification error rate in the experiments conducted on development section of ASVspoof $2015$ corpus. Even though we are able to accurately classify synthetic speech for five different spoofing attacks, it is to be noted that all the spoofing techniques tested here are known i.e. similar types of spoofed speech files are used for training. Therefore, it is interesting to study the performance of the proposed features in presence of unknown spoofing attacks. In future, experiments will be conducted on the evaluation part of the database which contains additional test data as well as data from unseen spoofing attacks.

\section*{Acknowledgement}
This work is partially supported by Indian Space Research Organization (ISRO), Government of India. We would also like to thank Dr. Md Sahidullah from University of Eastern Finland, Finland for his valuable suggestions.

\bibliographystyle{ieeetran}
\bibliography{latexbib}

\end{document}